\documentclass[preprint, numbers]{sigplanconf}

\pdfoutput=1

\usepackage[hyperindex,breaklinks]{hyperref}
\usepackage{amsmath}
\usepackage{cleveref}
\usepackage{booktabs}
\usepackage{fancyvrb}
\usepackage{color}
\usepackage{array}
\usepackage{fmtcount}
\usepackage{draftwatermark}

\SetWatermarkLightness{0.9}
\SetWatermarkText{DRAFT}

\newcommand{\final}[1]{#1}
\newcommand{\xxx}{LifeJacket}

\makeatletter
\newcolumntype{V}[1]{>{\topsep=0pt\@minipagetrue}p{#1}<{\vspace{-\baselineskip}}}
\makeatother

\newcommand{\numtranslated}{54}
\newcommand{\numverified}{43}
\newcommand{\numbugs}{8}
\newcommand{\numnewbugs}{3}
\newcommand{\numtimeouts}{4}
\newcommand{\numreported}{4}

\begin{document}

\title{\xxx{}: Verifying precise floating-point
optimizations in LLVM}

\authorinfo{Andres N\"otzli\and Fraser Brown}
           {Stanford University}
           {\{noetzli,mlfbrown\}@stanford.edu}

\maketitle

\begin{abstract}
  Optimizing floating-point arithmetic is vital because it is ubiquitous,
  costly, and used in compute-heavy workloads. Implementing precise
  optimizations correctly, however, is difficult, since developers must account
  for all the esoteric properties of floating-point arithmetic to ensure that
  their transformations do not alter the output of a program. Manual reasoning
  is error prone and stifles incorporation of new optimizations.

  We present an approach to automate reasoning about floating-point
  optimizations using satisfiability modulo theories (SMT) solvers.  We
  implement the approach in \xxx{}, a system for automatically verifying
  precise floating-point optimizations for the LLVM assembly language. We have
  used \xxx{} to verify \numverified{} LLVM optimizations and to discover
  \numberstringnum{\numbugs} incorrect ones, including
  \numberstringnum{\numnewbugs} previously unreported problems. \xxx{} is an
  open source extension of the Alive system for optimization verification.

\end{abstract}

\section{Introduction}
In this paper, we present \xxx{}, a system for automatically verifying
floating-point optimizations. Floating-point arithmetic is ubiquitous---modern
hardware architectures natively support it and programming languages treat it
as a canonical representation of real numbers---but writing correct
floating-point programs is difficult. Optimizing these programs is even more
difficult. Unfortunately, despite hardware support, floating-point computations
are still expensive, so avoiding optimization is undesirable.

Reasoning about floating-point optimizations and programs is difficult because
of floating-point arithmetic's unintuitive semantics. Floating-point arithmetic
is inherently imprecise and lossy, and programmers must account for rounding,
signed zeroes, special values, and non-associativity~\cite{goldberg1991every}.
Before the standardization, a wide range of incompatible floating-point
hardware with varying support for range, precision, and rounding existed. These
implementations were not only incompatible but also had undesirable properties
such as numbers that were not equal to zero for comparisons but were treated as
zeros for multiplication and division~\cite{severance1998interview}. The IEEE
754-1985 standard and its stricter IEEE 754-2008 successor were carefully
designed to avoid many of these pitfalls and designed for (contrary to popular
opinion, perhaps) non-expert users. Despite these advances, program correctness
and reproducibility still rests on a fragile interplay between developers,
programming languages, compilers, and hardware implementations.

Compiler optimizations that alter the semantics of programs, even in subtle
ways, can confuse users, make problems hard to debug, and cause cascading
issues. IEEE 754-2008 acknowledges this by recommending that language standards
and implementations provide means to generate \textit{reproducible} results for
programs, independent from optimizations. In practice, many transformations
that are valid for real numbers, change the precision of floating-point
expressions. As a result, compilers optimizing floating-point programs face the
dilemma of choosing between speed and reproducibility. They often address this
dilemma by dividing floating-point optimizations into two groups, precise and
imprecise optimizations, where imprecise optimizations are optional (e.g. the
\texttt{-ffast-math} flag in \texttt{clang}). While precise optimizations
always produce the same result, imprecise ones produce reasonable results on
common inputs (e.g. not for special values) but are arbitrarily bad in the
general case. To implement precise optimizations, developers have to reason
about all edge cases of floating-point arithmetic, making it challenging to
avoid bugs.

To illustrate the challenge of developing floating-point optimizations,
\Cref{fig:bug} shows an example of an invalid transformation implemented in
LLVM 3.7.1. We discuss the specification language in more detail in
\Cref{sec:alive} but, at a high-level, the transformation simplifies ${+0.0} -
({-0.0} - x)$ to $x$, an optimization that is correct in the realm of real
numbers. Because floating-point numbers distinguish between negative and
positive zero, however, the optimization is not valid if $x = {-0.0}$, because
the original code returns $+0.0$ and the optimized code returns ${-0.0}$. While
the zero's sign may be insignificant for many applications, the unexpected sign
change may cause a ripple effect. For example, the reciprocal of zero is
defined as $1 / {+0.0} = {+\infty}$ and $1 / {-0.0} = {-\infty}$.

Since reasoning manually about floating-point operations and optimizations is
difficult, we argue that automated reasoning can help ensure correct
optimizations. The goal of \xxx{} is to allow LLVM developers to automatically
verify precise floating-point optimizations. Our work focuses on precise
optimizations because they are both more amenable to verification and arguably
harder to get right. \xxx{} builds on Alive~\cite{lopes2015provably}, a tool
for verifying LLVM optimizations, extending it with floating-point support.

\begin{figure}
\small
\begin{Verbatim}
Name: PR26746
%a = fsub -0.0, %x
%r = fsub +0.0, %a
  =>
%r = %x
\end{Verbatim}

\caption{Incorrect transformation involving floating-point instructions in LLVM
3.7.1.}
\label{fig:bug}

\end{figure}

Our contributions are as follows:

\begin{itemize}

\item We describe the background for verifying precise floating-point
optimizations in LLVM and propose an approach using SMT solvers.

\item We implemented the approach in \xxx{}, an open source fork of Alive that
adds support for floating-point types, floating-point instructions,
floating-point predicates, and certain fast-math flags.

\item We validated the approach by verifying \numverified{} optimizations.
\xxx{} finds \numbugs{} incorrect optimizations, including
\numberstringnum{\numnewbugs} previously unreported problems in LLVM 3.7.1.

\end{itemize}

In addition to the core contributions, our work also lead to the discovery of
two issues in Z3~\cite{de2008z3}, the SMT solver used by \xxx{}, related to
floating-point support.

\section{Related Work} \label{sec:related}

Alive is a system that verifies LLVM peephole optimizations. \xxx{} is a fork
of this project that extends it with support for floating-point arithmetic.  We
are not the only ones interested in verifying floating-point optimizations;
close to the submission deadline, we found that one of the Alive authors had
independently begun a reimplementation of Alive that seems to include support
for floating-point
arithmetic.\footnote{\url{https://github.com/rutgers-apl/alive-nj}}

Our work intersects with the areas of compiler correctness, optimization
correctness, and analysing floating-point expressions.


Research on compiler correctness has addressed floating-point and
floating-point optimizations. CompCert, a formally-verified compiler, supports
IEEE 754-2008 floating-point types and implements two floating-point
optimizations~\cite{boldo2015verified}. In CompCert, developers use Coq to
prove optimizations correct, while \xxx{} proves optimization correctness
automatically.

Regarding optimization correctness, researchers have explored both the
consequences of existing optimizations and techniques for generating new
optimizations. Recent work has discussed consequences of unexpected
optimizations~\cite{wang2013towards}. In terms of new optimizations,
STOKE~\cite{schkufza2014stochastic} is a stochastic optimizer that supports
floating-point arithmetic and verifies instances of floating-point
optimizations with random testing. Souper~\cite{souper} discovers new LLVM
peephole optimizations using an SMT solver. Similarly, Optgen generates
peephole optimizations and verifies them using an SMT
solver~\cite{buchwald2015optgen}. All of these approaches are concerned with
the correctness of new optimizations, while our work focuses on existing ones.
Vellvm, a framework for verifying LLVM optimizations and transformations using
Coq, also operates on existing transformations but does not do automatic
reasoning.

Researchers have explored debugging floating-point
accuracy~\cite{chiang2014efficient} and improving the accuracy of
floating-point expressions~\cite{panchekha2015automatically}. These efforts are
more closely related to imprecise optimizations and provide techniques that
could be used to analyze them. Z3's support for reasoning about floating-point
arithmetic relies on a model construction procedure instead of naive
bit-blasting~\cite{zeljic2014approximations}.



\section{Background}

\begin{table*}
\small
\centering
\begin{tabular}{lp{9cm}V{6.5cm}}
\toprule
Flag & Description & Formula \\
\midrule
\texttt{nnan} &
Assume arguments and result are not \texttt{NaN}. Result undefined over
\texttt{NaN}s.
&
\begin{Verbatim}
ite (or (isNaN a) (isNaN b) (isNaN r)
    (x (_ FP <ebits> <sbits>)) r
\end{Verbatim}
\\
\texttt{ninf} &
Assume arguments and result are not $\pm\infty$. Result undefined over
$\pm\infty$.
&
\begin{Verbatim}
ite (or (isInf  a) (isInf b) (isInf r))
    (x (_ FP <ebits> <sbits>)) r
\end{Verbatim}
\\
\texttt{nsz} &
Allow optimizations to treat the sign of a zero argument or result as
insignificant.
&
\begin{Verbatim}
or (a = b) (and (isZero a) (isZero b))
\end{Verbatim}
\\
\bottomrule
\end{tabular}

\caption{Fast-math flags that \xxx{} supports. The \texttt{isNaN} and
\texttt{isInf} are not part of the SMT-LIB standard but supported in Z3's
Python interface and used for illustration purposes here. The variable
\texttt{x} is a fresh, unconstrained variable, \texttt{a} and \texttt{b} are
the SMT formulas of the operands, \texttt{r} of the result. The formula for
\texttt{nsz} replaces the standard equality check \texttt{a = b}.}

\label{tab:fast-math}
\end{table*}

\noindent Our work verifies LLVM floating-point optimizations. These optimizations
take place on LLVM assembly language, a human-readable, 
low-level language. The language serves as a common representation for
optimizations, transformations, and analyses. Front ends (like \texttt{clang})
output the language, and, later, back ends use it to generate machine code for
different architectures.

Our focus is verifying peephole optimizations implemented in LLVM's InstCombine
pass.  This pass replaces small subtrees in the program tree without changing
the control-flow graph. Alive already verifies some InstCombine optimizations,
but it does not support optimizations involving floating-point arithmetic.
Instead of building \xxx{} from scratch, we extends Alive with the machinery to
verify floating-point optimizations. To give the necessary context for
discussing our implementation in \Cref{sec:implementation}, we describe LLVM's
floating-point types and instructions and give a brief overview of Alive.

\subsection{Floating-point arithmetic in LLVM} \label{sec:fpllvm}

In the following, we discuss LLVM's semantics of floating-point types and
instructions. The information is largely based on the LLVM Language Reference
Manual for LLVM 3.7.1~\cite{llvm-lang-ref} and the IEEE 754-2008 standard. For
completeness, we note that the language reference does not explicitly state
that LLVM floating-point arithmetic is based on IEEE 754. However, the language
reference refers to the IEEE standard multiple times, and LLVM's floating-point software
implementation \texttt{APFloat} is explicitly based on the standard.

\paragraph{Floating-point types} LLVM defines six different floating-point
types with bit-widths ranging from 16 bit to 128 bit. Floating-point values are
stored in the IEEE binary interchange format, which encodes them in three
parts: the sign $s$, the exponent $e$ and the significand $t$. The value of a
normal floating-point number is given by: $(-1)^s \times (1 + 2^{1-p} \times t)
\times 2^{e - bias}$, where $bias = 2^{w - 1} - 1$ and $w$ is the number of
bits in the exponent. The range of the exponents for normal floating-point
numbers is $[1, 2^w-2]$. Exponents outside of this range are used to encode
special values: subnormal numbers, Not-a-Number values (\texttt{NaN}s), and
infinities.

Floating-point zeros are signed, meaning that $-0.0$ and $+0.0$ are distinct.
While most operations ignore the sign of a zero, the sign has an observable
effect in some situations: a division by zero (generally) returns $+\infty$ or
$-\infty$ depending on the zero's sign, for example. As a consequence, $x = y$
does not imply $\frac{1}{x} = \frac{1}{y}$. If $x = 0$ and $y = -0$, $x = y$ is
true, since floating point $0 = -0$. On the other hand, $\frac{1}{x} =
\frac{1}{y}$ is false, since $\frac{1}{0} = \infty \neq -\infty =
\frac{1}{-0}$.

Infinities ($\pm\infty$) are used to represent an overflow or a division by
zero. They are encoded by setting $t = 0$ and $e = 2^w - 1$. Subnormal numbers,
on the other hand, are numbers with exponents below the minimum exponent;
normal floating-point numbers have an implicit leading $1$ in the significand
that prevents them from representing these numbers.  The IEEE standard defines
the value for subnormal numbers as: $(-1)^s \times (0 + 2^{1-p} \times t)
\times 2^{e_{min}}$, where $e_{min} = 1 - bias$.

\texttt{NaN}s are used to represent the result of an invalid operation (such as
$\infty - \infty$) and are described by $e = 2^w - 1$ and a non-zero $t$.
There are two types of \texttt{NaN}s: quiet \texttt{NaN}s (\texttt{qNaN}s) and
signalling \texttt{NaN}s (\texttt{sNaN}s). The first bit in the significand
determines the type of \texttt{NaN} ($1$ in the case of a \texttt{qNaN}) and
the remaining bits can be used to encode debug information. Operations
generally propagate \texttt{qNaN}s and quiet \texttt{sNaN}s: If one of the
operands is \texttt{qNaN}, the result is \texttt{qNaN}, if the operand is an
\texttt{sNaN}, it is quieted by setting the first bit to $1$.

Floating-point exceptions occur in situations like division by zero or
computation involving an \texttt{sNaN}. By default, floating-point exceptions
do not alter control-flow but raise a status flag and return a default result
(e.g. a \texttt{qNaN}).

\paragraph{Floating-point instructions} In its assembly language, LLVM defines
several instructions for binary floating-point operations (\texttt{fadd},
\texttt{fsub}, \texttt{fmul}, \texttt{fdiv},~\ldots), conversion instructions
(\texttt{fptrunc}, \texttt{fpext}, \texttt{fptoui}, \texttt{uitofp},~\ldots),
and allows floating-point arguments in other operations (e.g. \texttt{select}).
We assert that floating-point instructions cannot generate poison values
(values that cause undefined behavior for instructions that depend on them) or
result in undefined behavior. The documentation is not entirely clear but our
interpretation is that undefined behavior does not occur in the absence of
\texttt{sNaN}s and that \texttt{sNaN}s are not fully supported.

While IEEE 754-2008 defines different rounding modes, LLVM does not yet allow
users to specify them. As a consequence, the rounding performed by
\texttt{fptrunc} (casting a floating-point value to a smaller floating-point
type) is undefined for inexact results.

\paragraph{Fast-math flags}
Some programs either do not depend on the exact
semantics of special floating-point values or do not expect special values
(such as \texttt{NaN}) to occur. To specify these cases, LLVM binary operators can
provide \emph{fast-math flags}, which allow LLVM to do additional optimizations with
the knowledge that special values will not occur. \Cref{tab:fast-math}
summarizes the fast-math flags that \xxx{} supports. There are two additional
flags, \texttt{arcp} (allows replacing arguments of a division with the
reciprocal) and \texttt{fast} (allows imprecise optimizations), that we do not
support.

\paragraph{Discussion} The properties of floating-point arithmetic discussed in
this section hint at how difficult it is to manually reason about
floating-point optimizations. The floating-point standard is complex, so
compilers do not always follow it completely---as we mentioned earlier, LLVM
does not currently support different rounding modes.\footnote{More details:
\url{http://lists.llvm.org/pipermail/llvm-dev/2016-February/094869.html}.}
Similarly, it does not yet support access to the floating-point environment,
which makes reliable checks for floating-point exceptions in \texttt{clang}
impossible, for example. This runs counter to the IEEE standard, which defines
reproducability as including ``invalid operation,'' ``division by zero,'' and
``overflow'' exceptions.

\subsection{Verifying transformations with Alive} \label{sec:alive}
Alive is a tool that verifies peephole optimizations on LLVM's intermediate
representation; these optimizations are expressed (as input) in a
domain-specific language. 
At a high level, verifying an optimization with Alive takes the following steps:

\begin{enumerate}

\item The user specifies a new or an existing LLVM optimization using the Alive
language.

\item Alive translates the optimization into a series of SMT queries that
express the equivalence of the source and the target.

\item Alive uses Z3, an SMT solver, to check whether any combination of values
makes the source and target disagree. If the optimization is incorrect, Alive
returns a counter-example that breaks the optimization.

\end{enumerate}

Alive specializes in peephole optimizations that are highly local and do not
alter the control-flow graph of a program. This type of optimization is
performed by the LLVM InstCombine pass in \texttt{lib/Transforms/InstCombine}
and InstructionSimplify in \texttt{lib/Analysis}.

Alive can also generate code for an optimizer pass that performs all of the
verified optimizations. We do not discuss this feature further since \xxx{}
does not support it for floating-point optimizations. In the following, we
discuss the Alive language and the role of SMT solvers in proving optimization
correctness.

\paragraph{Specifying transformations with the Alive language} In the
domain-specific Alive language, each transformation consists of a list of
preconditions, a source template, and a target template. Alive verifies whether
it is safe to replace the instructions in the source template with the
instructions in the target given that the preconditions hold. 
\Cref{fig:bug} is an example of a transformation in the Alive language. This
transformation has no preconditions, so it always applies. The instructions
above the ``\texttt{=>}'' delimiter are the source template, while the target
template are below.

Preconditions are logical expressions enforced by the compiler at compile-time
and Alive takes them for granted. The precondition \texttt{isNormal(\%x)}, for
example, expresses the fact that an optimization only holds when \texttt{\%x}
is a normal floating-point value.

Alive interprets the instructions in the sources and targets as expression
trees, so the order of instructions does not matter, only the dependencies.
Verifying the equivalence of the source and the target is done on the root of
the trees. The arguments for instructions are either inputs (e.g.
\texttt{\%x}), constant expressions (e.g. \texttt{C}), or immediate values
(e.g. \texttt{0.0}). Inputs model registers, constant expressions correspond to
computations that LLVM performs at compile-time, and immediate values are
values known at verification time. Constant expressions consist of constant
functions and compile-time constants. Inputs and constant expressions can be
subjects for predicates in the precondition.

In contrast to actual LLVM code, the Alive language does not require type
information for instructions and inputs. Instead, it uses the types expected by
instructions to restrict types and bit-widths of types. Then, it issues an SMT
query that encodes these constraints to infer all possible types and sizes of
registers, constants, and values. This mirrors the fact that LLVM optimizations
often apply to multiple bit-widths and makes specifying optimizations less
repetitive. Alive instantiates the source and target templates with the
possible type and size combinations and verifies each instance.

\begin{figure}
\small
\centering
\begin{minipage}{3.5cm}
Incorrect:
\begin{Verbatim}
%r = fdiv %x, undef
  =>
%r = undef
\end{Verbatim}
\end{minipage} %
\begin{minipage}{3.5cm}
Correct:
\begin{Verbatim}
%r = fdiv %x, undef
  =>
%r = NaN
\end{Verbatim}
\end{minipage} %

\caption{Example of a problematic optimization using \texttt{undef} on the left
and a better version on the right. If \texttt{\%x} is \texttt{NaN} then
\texttt{\%r} can only be \texttt{NaN}, so \texttt{\%r} cannot be
\texttt{undef}.}

\label{fig:undef}

\end{figure}

Undefined values (\texttt{undef}) in LLVM represent input values of arbitrary
bit-patterns when used and may be of any type. For each \texttt{undef} value in
the target template, Alive has to verify that any value can be produced and for
each \texttt{undef} value in the source, Alive may assume any convenient value.
\Cref{fig:undef} is a known incorrect optimization in LLVM that \xxx{} confirms
and that illustrates this concept: The source template cannot produce all
possible bit-patterns, so it cannot be replaced with
\texttt{undef}.\footnote{Discussion on this optimization:
\url{https://groups.google.com/d/topic/llvm-dev/iRb0gxroT9o/discussion}}

\paragraph{Verifying transformations with SMT solvers}
Alive translates the source and target template into SMT formulas.
For each possible combination of variable types in the templates,
it creates SMT formulas for definedness constraints, poison-free constraints,
and the execution values for the source and target. Alive checks
the definedness and poison-free constraints of the source and target for
consistency. These checks are not directly relevant to floating-point
arithmetic, so we do not discuss them further. Instead, we deal more directly with
the execution values of the source and target. 

An optimization is only correct if the source and the target always produce the
same value. To check this property, Alive asks an SMT solver to verify that
$\texttt{preconditions} \wedge \texttt{src\_formula} \ne \texttt{tgt\_formula}$
is unsatisfiable---that there is no assignment that can make the formula true.
If there is, the optimization is \textit{incorrect}: there is an assignment for
which the source value is different from the target value. When Alive
encounters an incorrect optimization, it uses the output of the SMT solver to
return a counterexample in the form of input and constant assignments that lead
to different source and target values.


Ultimately, Alive relies on Z3 to determine whether an optimization is correct
(by answering the SMT queries). \xxx{} would have been impossible without Z3's
floating-point support, which was added in version 4.4.0 by implementing the
SMT-LIB standard for floating-point arithmetic~\cite{smtFPA2010} less than a
year ago.


\section{Implementation} \label{sec:implementation}
Our implementation extends Alive in four major ways: It adds support for
floating-point types, floating-point instructions, floating-point predicates,
and fast-math flags. In the following, we describe our work in those areas,
briefly comment on our experience with floating-point support in Z3, and
conclude with a discussion of the limitations of the current version of \xxx{}.

\paragraph{Floating-point types}
\xxx{} implements support for \texttt{half}, \texttt{single}, and
\texttt{double} floating-points.
Alive itself provides support for integer and pointer types of arbitrary
bit-widths up to 64 bit. Following the philosophy of the original implementation,
we do not require users to explicitly annotate floating-point types. Instead,
we use a logical disjunction (in the SMT formula for type constraints)
to limit floating-point types to bit-widths of 16, 32, or 64 bits. Then, we use
Alive's existing mechanisms to determine all possible type combinations for each
optimization (as discussed in \Cref{sec:alive}). 

Adding a new type required us to relax some assumptions, e.g. that the
arguments of \texttt{select} are integers. Additionally, we modified the parser
to support floating-point immediate values.

\paragraph{Floating-point predicates and constant functions} \xxx{} adds
precondition predicates and constant functions related to floating-point
arithmetic.

Recall that preconditions are logical formulas that describe facts that must be
true in order to perform an optimization; they are fulfilled by LLVM and
assumed by Alive. In the context of floating-point optimizations, preconditions
may include predicates about the type of a floating-point number (e.g.
\texttt{isNormal(\%x)} to make sure that \texttt{\%x} is a normal
floating-point number) or checks to ensure that conversions are lossless. We
discuss more predicates in the following paragraphs.


Constant functions mirror computation performed by LLVM at compile-time and are
evaluated by Alive symbolically at verification-time.  For example, the
constant function \texttt{fptosi(C)} (not to be confused with the instruction)
converts a floating point number to a signed integer, corresponding to a
conversion LLVM does at compile time. Constant expressions (expressions that
contain constant functions) can be assigned to registers in the target
template, mirroring the common strategy of optimizing operations by partially
evaluating them at compile-time.

In contrast to Alive, \xxx{} supports precondition predicates that refer to
constant expressions in target templates. For example, some optimizations have
restrictions about precise conversions, and we express those restrictions in
the precondition. If the target converts a floating-point constant to an
integer with \texttt{\%c = fptosi(C)}, then the precondition can ensure that
the conversion is lossless by including \texttt{sitofp(\%c) == C} (which
guarantees that converting the number back and forth results in the original
number).  If the precondition does not refer to \texttt{\%c} in the target and
instead imposes \texttt{sitofp(fptosi(C)) == C} then it would not restrict the
bit-width of \texttt{\%c}, so \texttt{\%c} could be too narrow to represent the
number.

%

\paragraph{Floating-point instructions}
Our implementation supports binary floating-point instructions (\texttt{fadd},
\texttt{fsub}, \texttt{fmul}, \texttt{fdiv}, and \texttt{frem}), conversions
involving floating-point numbers (\texttt{fptrunc}, \texttt{fpext},
\texttt{fptoui}, \texttt{fptosi}, \texttt{uitofp}, \texttt{sitofp}), the
\texttt{fabs} intrinsic, and floating-point comparisons (\texttt{fcmp}). Most
of these instructions directly correspond to operations
that the SMT-LIB for floating-point standard supports, so translating them to
SMT formulas is straightforward.  Next, we discuss our support for
\texttt{frem}, \texttt{fcmp}, conversions, and the equivalence check for
floating-point optimizations.

%

\begin{figure}
\small

\begin{Verbatim}
double fmod(double x, double y) {
  double result;
  result = remainder(fabs(x), (y = fabs(y)));
  if (signbit(result)) result += y;
  return copysign(result, x);
}
\end{Verbatim}
\vspace{-0.2cm}
\noindent\rule{\columnwidth}{0.4pt}
\begin{Verbatim}
(= abs_y (abs y))
(= r (remainder (abs x) abs_y))
(= r' (ite (isNeg r) (+ RNE r abs_y) r))
(= fmod (ite (xor (isNeg x) (isNeg r')) (- r') r))
\end{Verbatim}

\caption{The \texttt{fmod} function implemented using IEEE \texttt{remainder}
as suggested by the C standard and an informal representation of the implementation used by \xxx{}.}

\label{fig:fmod}
\end{figure}

The \texttt{frem} instruction does not correspond to \texttt{remainder} as
defined by IEEE 754 but rather to \texttt{fmod} in the C POSIX library, so
translating it to an SMT formula involves multiple operations.  Both
\texttt{fmod} and \texttt{remainder} calculate $x - n * y$ (where $n$ is
$\frac{x}{y}$), but \texttt{fmod} rounds toward zero whereas \texttt{remainder}
rounds to the nearest value and ties to even. \Cref{fig:fmod} shows how the C
standard defines \texttt{fmod} in terms of \texttt{remainder} for
\texttt{double}s~\cite[\S F.10.7.1]{c11} and the corresponding SMT formula that
\xxx{} implements. The formula uses a fixed rounding-mode because the
rounding-mode of the environment does not affect \texttt{fmod}.

The \texttt{fcmp} instruction compares two floating-point values. In addition
to the two floating-point values, it expects a third operand, the
\emph{condition code}. The condition code determines the type of comparison.
There are two larger genres of comparison: \emph{ordered} comparisons can only
be true if none of the inputs are \texttt{NaN} and \emph{unordered} comparisons
are true if any of the inputs is \texttt{NaN}. LLVM supports an ordered version
and an unordered version of the usual comparisons such as equality, inequality,
greater-than, etc. Additionally, there are condition codes that just check
whether both inputs are not \texttt{NaN} (\texttt{ord}) or any of the inputs
are \texttt{NaN} (\texttt{uno}).

Optimizations involving comparisons often apply to multiple condition codes. To
allow users to efficiently describe such optimizations, \xxx{} supports
predicates in the precondition that describe the applicable set of condition
codes. For example, there are predicates for constraining the set of condition
codes to either ordered or unordered conditions.  We also support predicates
that express a relationship between multiple condition codes. This is useful,
for example, to describe an optimization that performs a multiplication by
negative one on both sides: To replace the comparison (\texttt{C1}) between
\texttt{-x} and \texttt{C} with the comparison (\texttt{C2}) between \texttt{x}
and \texttt{-C}, we use the \texttt{swap(C1, C2)} predicate.

When no sensible conversion between floating-point values and integers is
possible, LLVM defaults to returning \texttt{undef}. For conversions from
floating-point to integer value (signed or unsigned), \xxx{} checks whether the
(symbolic) floating-point value is \texttt{NaN}, $\pm\infty$, too small, or too
large and returns \texttt{undef} if necessary. Conversions from integer to
floating-point values, similarly return \texttt{undef} for values that are too
small or too large.



Recall that \xxx{} must determine the unsatisfiability of
$\texttt{precondition} \wedge \texttt{src\_formula} \ne \texttt{tgt\_formula}$
to verify optimizations. The SMT-LIB standard defines two equality operators
for floating-point, one implementing bit-wise equality, and one implementing
the IEEE equality operator. The latter operator treats signed zeros as
equal and \texttt{NaN}s as different, so using it to verify optimizations would
not work, since it would accept optimizations that produce different zeros and
reject source-target pairs that both produce \texttt{NaN}. The bit-wise
equality works, because SMT-LIB uses a single \texttt{NaN} value (recall that
there are multiple bit-patterns that correspond to \texttt{NaN}). While this
is convenient, it also means that we cannot model different \texttt{NaN}s. We
discuss the implications later.

\paragraph{Fast-math flags}
\xxx{} currently supports three of the five fast-math flags that LLVM
implements: \texttt{nnan}, \texttt{ninf}, and
\texttt{nsz}.

\xxx{} handles the \texttt{nnan} and \texttt{ninf} flags in a similar way by
modifying the SMT formula for the instruction on which the flag appears. As
\Cref{tab:fast-math} shows, if the instruction's arguments or result is a
\texttt{NaN} or $\pm \infty$, respectively, the formula returns a fresh
unconstrained variable that it treats as an \texttt{undef} value. This is a
direct translation from the description in the language reference and works for
root and non-root instructions.

The \texttt{nsz} flag is different: Instead of relaxing the requirements for
the behavior for certain inputs and results, it states that the sign of a zero
value can be ignored. This primarily affects how \xxx{} compares the source and
target values: it adds a logical conjunction to the SMT query that states that
the source and target values are only different if both are nonzero (shown in
\Cref{tab:fast-math}). The flag itself has no effect on zero values at runtime,
meaning that it does not affect the computation performed by instructions with
the flag. Thus, we do not change the SMT formula for the instruction.

Since the \texttt{nsz} flag has no direct effect on how LLVM does matching,
this flag \emph{also} does not change the significance of the sign of immediate
zeros (e.g. \texttt{+0.0}) in the optimization templates. Instead, we mirror
how LLVM determines whether an optimization applies. In LLVM, optimizations
that match a certain sign of zero do not automatically apply to other zeros
when the \texttt{nsz} flag is set. For example, an optimization that applies to
\texttt{fadd x, -0.0} does \emph{not} automatically apply to \texttt{fadd nsz
x, +0.0}. If applicable, developers explicitly match any zero if the
\texttt{nsz} flag is set.  We mirror this design by implementing an
\texttt{AnyZero(C)} predicate, which makes \texttt{C} negative or positive
zero.

\begin{figure*}
\centering
\small

\begin{minipage}[t]{5cm}
\begin{Verbatim}
Name: PR26958
Precondition: AnyZero(C0)
%a = fsub nnan ninf C0, %x
%r = fadd %x, %a
  =>
%r = 0.0
\end{Verbatim}
\end{minipage} %
\begin{minipage}[t]{5cm}
\begin{Verbatim}
Name: PR26943
%a = select i1 %c, 0.0, C
%r = frem %x, %a
  =>
%r = frem %x, C
\end{Verbatim}
\end{minipage} %
\begin{minipage}[t]{5cm}
\begin{Verbatim}
Name: PR27036
Precondition: hasOneUse(%a) &&
  hasOneUse(%b) &&
  WillNotOverflowSignedAdd(%x, %y)
%a = sitofp %x
%b = sitofp %y
%r = fadd %a, %b
  =>
%c = add nsw %x, %y
%r = sitofp %c
\end{Verbatim}
\end{minipage}

\caption{New bugs in LLVM 3.7.1 found by \xxx{}.}
\label{fig:bugs}
\end{figure*}

\paragraph{Limitations} While \Cref{sec:evaluation} shows that \xxx{} is a
useful tool, it does not support all floating-point types and imprecise
optimizations, uses a fixed rounding-mode, and does not model floating-point
exceptions and debug information in \texttt{NaN}s.

Currently, \xxx{} does not support LLVM's vectors and
the two 128-bit and the 80-bit floating-point types. Supporting those would
likely not require fundamental changes.

There are many imprecise optimizations in LLVM. These optimizations need a
different style of verification because they do not make any guarantees about
how much they affect the program output. A possible way to deal with these
optimizations would be to verify that they are correct for real numbers and
estimate accuracy changes by randomly sampling inputs,
similar to Herbie~\cite{panchekha2015automatically}.

\xxx{}'s verification ultimately relies on the SMT-LIB standard for
floating-point arithmetic. The standard corresponds to IEEE 754-2008
but it only defines a single \texttt{NaN} value and does
not distinguish between signalling and quiet \texttt{NaN}s. Thus,
our implementation cannot verify whether an operation with \texttt{NaN}
operands returns one of the input \texttt{NaN}s, propagating debug information
encoded in the \texttt{NaN}, as recommended by the IEEE standard. In practice,
LLVM does not attempt to preserve information in \texttt{NaN}s, so this
limitation does not affect our ability to verify LLVM optimizations.  We do not
model floating-point exceptions, either, since LLVM does not currently make
guarantees about handling floating-point exceptions. Floating-point exceptions
could be verified with separate SMT queries, similar to how Alive verifies
definedness.

\xxx{} currently rounds to nearest and ties to the nearest even digit,
mirroring the most common rounding-mode. Even though LLVM does not yet support
different rounding-modes, we are planning to add support soon.

The limited type and rounding-mode support and missing floating-point
exceptions make our implementation unsound at worst: \xxx{} may label some
incorrect optimizations as correct, but optimizations labelled as incorrect are
certainly wrong.

\paragraph{Working with Z3} Even though Z3's implementation of floating-point
support is recent, we found it to be an effective tool for the job. Due to the
youth of the floating-point support, we found that  \xxx{} does not work with
the newest release of Z3 because of issues in the implementation and the Python
API. During the development of \xxx{}, we reported issues that were fixed
quickly and fixed some issues, mostly in the Python API, ourselves. This
suggests that \xxx{} is an interesting test case for floating-point support in
SMT solvers.

\section{Evaluation} \label{sec:evaluation}

To evaluate \xxx{}, we translated \numtranslated{} optimizations from LLVM
3.7.1 into the Alive language and tried to verify them. We discovered
\numbugs{} incorrect optimizations and verified \numverified{} optimizations to
be correct. In the following, we outline the optimizations that we checked and
describe the bugs that we found.

We performed our evaluation on a machine with an Intel i3-4160 CPU and 8 GB
RAM, running Ubuntu 15.10. We compiled Z3 commit \texttt{b66fc4e}\footnote{Full
disclaimer: We ran into regression issues with this version, we verified some
optimizations with an older version, will change for camera ready.} with GCC
5.2.1, the default compiler, used the \texttt{qffpbv} tactic, and chose a 5
minute timeout for SMT queries. \Cref{fig:opts} summarizes the results for the
different source files: AddSub contains optimizations with
\texttt{fadd}/\texttt{fsub} at the root, MulDivRem with
\texttt{fmul}/\texttt{fdiv}/\texttt{frem}, Compares deals with \texttt{fcmp}s
and Simplify contains simple optimizations for all instructions.

Using this process, \xxx{} found \numverified{} out of \numtranslated{}
optimizations to be correct. \xxx{} timed out on \numberstringnum{\numtimeouts}
optimizations. The AddSub optimization that times out contains a
\texttt{sitofp} instruction and verification is slow for integers with a large
bit-width. The two MulDivRem optimizations that timeout both contain
\texttt{nsz} flags and \texttt{AnyZero} predicates. Similar optimizations
without those features do not timeout. In general, \texttt{fdiv} seems to slow
down verification as seems to be the case for the timeout in Simplify. Out of
the \numbugs{} optimizations that we found to be incorrect,
\numberstringnum{\numreported} had been reported.  The bug in \Cref{fig:bug}
had already been fixed in a newer version of LLVM when we discovered it.  The
rest of the reported bugs resembled the example in \Cref{fig:undef} and are all
caused by an unjustified \texttt{undef} in the target. \Cref{fig:bugs} depicts
the \numberstringnum{\numnewbugs} previously unreported incorrect optimizations
that we reported to the LLVM developers.  We discuss these bugs in the next
paragraphs.

\texttt{PR26958} optimizes $(0 - x) + x$ to $0$. The implementation of this
optimization requires that the \texttt{nnan} and the \texttt{ninf} flag each
appear at least once on the source instructions. We translate four variants of
this instruction: One where both flags are on \texttt{fsub}, one where both are
on \texttt{fadd} and two where each instruction has one of the flags.  As it
turns out, it is not enough to have both flags on either of the instructions.
For the case where both flags are on \texttt{fsub}, the transformation is
invalid if \texttt{\%x} is \texttt{NaN} or $\pm \infty$. The \texttt{nnan} and
\texttt{ninf} flags require the optimized program to retain defined behavior
over \texttt{NaN} and $\pm \infty$, so \texttt{\%r} must be \texttt{0.0} even
for those inputs (if they resulted in undefined behavior, any result would be
correct). If \texttt{\%x} is \texttt{NaN}, however, then there is no value for
\texttt{\%a} that would result in \texttt{\%r} being \texttt{0.0} because
\texttt{NaN} added to any other number is \texttt{NaN}.

\texttt{PR26958} optimizes \texttt{fmod(x, c ? 0 : C)} to \texttt{fmod(x, C)}
(\texttt{select} acts like a ternary and \texttt{frem} corresponds to
\texttt{fmod}). The implementation of this optimization shares its code with
the same optimization for the \texttt{rem} instruction that deals with
integers. For integers, \texttt{rem \%x, 0} results in undefined behavior, so
the optimization is valid.  The POSIX standard specifies that \texttt{fmod(x,
0.0)} returns \texttt{NaN}, though, so the optimization is incorrect for
\texttt{frem} because \texttt{\%r} must be \texttt{NaN} and not \texttt{frem
\%x, C} if \texttt{\%a} is \texttt{0.0}.

\texttt{PR27036} illustrates the last incorrect optimization that \xxx{}
identified.  It transforms \texttt{(float) x + (float) y} into \texttt{(float)
(x + y)}, replacing an \texttt{fadd} instruction with a more efficient
\texttt{add}.  This transformation is invalid, though, since adding two rounded
numbers is not equivalent to adding two numbers and rounding the result. For
example, assuming 16-bit floating-point numbers, let \texttt{\%x = -4095} and
\texttt{\%y = 17}. In the portion of the source formula \texttt{\%a = sitofp
\%a}, \texttt{\%a} cannot store an exact number and stores \texttt{-4094}
instead.  The target formula, though, can accurately represent the result
\texttt{-4078} of the addition.

Our results confirm that it is difficult to write correct floating-point
optimizations; we found bugs in almost all the LLVM files from which we
collected our optimizations. Unsurprisingly, all of these bugs relate to
floating-point specific properties such as rounding, \texttt{NaN}, $\pm \infty$
inputs, and signed zeros. These edge cases are clearly difficult for
programmers to reason about.

\begin{table}
\centering
\small
\begin{tabular}{lrrr}
\toprule
File & Verified & Timeouts & Bugs \\
\midrule
AddSub & 7 & 1 & \final{1} \\
MulDivRem & \final{3} & \final{2} & \final{1} \\
Compares & \final{11} & \final{0} & \final{0} \\
Simplify & \final{22} & \final{1} & \final{6} \\
\midrule
Total & \numverified{} & 4 & \numbugs{} \\
\bottomrule
\end{tabular}

\caption{Number of optimizations verified, timeouts, and bugs.}
\label{fig:opts}

\end{table}






\section{Conclusion} In an ideal world, programming languages and compilers are
boring. They do what the user expects. They exhibit the same behavior with and
without optimization, at all optimization levels, and on all hardware.
``Boring,'' however, is surprisingly difficult to achieve, especially in the
context of the complicated semantics of floating-point arithmetic. With \xxx{},
we hope to make LLVM's precise floating-point optimizations more predictable
(and boring) by automatically checking them for correctness.

\bibliographystyle{plain}
\bibliography{paper.bib}

\end{document}